\documentclass[twoside,twocolumn,english]{revtex4-1}
\usepackage[T1]{fontenc}
\usepackage[latin9]{inputenc}
\usepackage{babel}
\usepackage{amsmath}
\usepackage{amssymb}
\usepackage{graphicx}
\usepackage{esint}
\usepackage[unicode=true,pdfusetitle,
 bookmarks=true,bookmarksnumbered=false,bookmarksopen=false,
 breaklinks=false,pdfborder={0 0 1},backref=false,colorlinks=false]
 {hyperref}

\makeatletter

\newcommand{\noun}[1]{\textsc{#1}}

\makeatother

\begin{document}
\title{Inability of a graph neural network heuristic to outperform greedy
algorithms in solving combinatorial optimization problems like Max-Cut}
\author{Stefan Boettcher}
\affiliation{Department of Physics, Emory University, Atlanta, GA 30322; USA}

\maketitle
\bigskip{}
\bigskip{}

\noun{Matters Arising} from Martin J. A. Schuetz et al. \emph{Nature
Machine Intelligence} https://doi.org/10.1038/s42256-022-00468-6 (2022).\bigskip{}

In Ref. \citep{Schuetz22}, Schuetz et al provide a scheme to employ
graph neural networks (GNN) as a heuristic to solve a variety of classical,
NP-hard combinatorial optimization problems. It describes how the
network is trained on sample instances and the resulting GNN heuristic
is evaluated applying widely used techniques to determine its ability
to succeed. Clearly, the idea of harnessing the powerful abilities
of such networks to ``learn'' the intricacies of complex, multimodal
energy landscapes in such a hands-off approach seems enticing. And
based on the observed performance, the heuristic promises to be highly
scalable, with a computational cost linear in the input size $n$,
although there is likely a significant overhead in the pre-factor
due to the GNN itself. However, closer inspection shows that the reported
results for this GNN are only minutely better than those for gradient
descent and get outperformed by a greedy algorithm, for example, for
Max-Cut. The discussion also highlights what I believe are some common
misconceptions in the evaluations of heuristics. 

\begin{figure}
\hfill{}\includegraphics[viewport=130bp 30bp 670bp 618bp,clip,width=1\columnwidth]{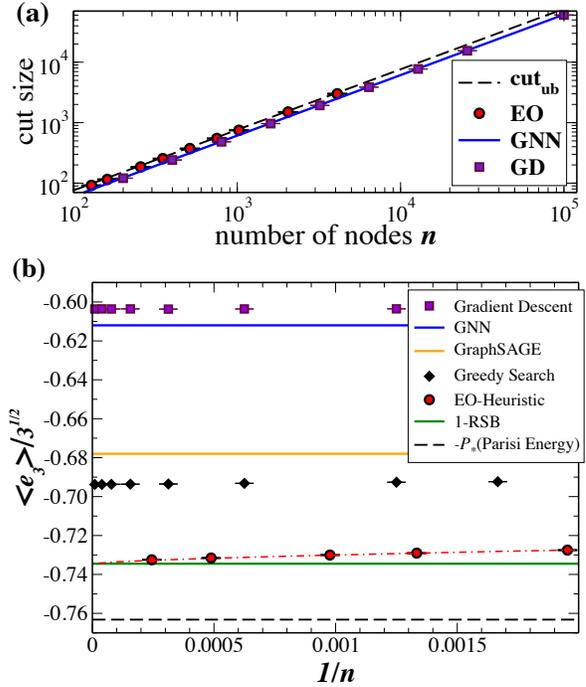}\hfill{}

\caption{\label{fig:cutsize}Results discussed in the text for various heuristics
and bounds for the Max-Cut problem on a 3-regular random graph ensemble,
(a) plotted for the raw cut-size as a function of problem size $n$,
and (b) as an extrapolation plot according to Eq. (\ref{eq:FSC}).
Note that in (b), a fit (red-dashed line) to the EO-data (circles)
suggests a non-linear asymptotic correction with $\sim1/n^{2/3}$
\citep{Boettcher03a}.}
\end{figure}

Among a variety of QUBO problems Ref. \citep{Schuetz22} consider
in their numerical evaluation of their GNN, I want to focus the discussion
here on Max-Cut. As explained in the context of Eq. (7), it is derived
from an Ising spin-glass Hamiltonian on a $d$-regular random graph
\footnote{Technically, their Hamiltonian in Eq. (7) pertains to an antiferromagnet
instead of a spin glass, but on such random graphs, both are equivalent
\citep{Zdeborova10}. } for $d=3$. (In the physics literature, for historical reason such
a graph is often referred to as a Bethe-lattice \citep{Mezard03,Boettcher03a}.)
Minimizing the energy of the Hamiltonian, $H$, maximizes the cut-size
$cut=-H$. The $cut$ results for the GNN (for both, $d=3$ and 5)
are presented in Fig. 4 of Ref. \citep{Schuetz22}, where they find
$cut\sim\gamma_{3}n$ with $\gamma_{3}\approx1.28$ via an asymptotic
fit to the GNN data obtained from averaging over randomly generated
instances of the problem for a progression of different problem sizes
$n$. In Fig. \ref{fig:cutsize}(a) here, I have recreated their Fig.
4, based on the value of $\gamma_{3}$ reported for GNN (blue line).
Like in Ref. \citep{Schuetz22}, I have also included what they describe
as a rigorous upper bound, $cut_{{\rm ub}}$ (black-dashed line),
which derives from an exact result obtained when $d=\infty$ \citep{parisi:80a}.
While the GNN results appear impressively close to that upper bound,
however, including two other sets of data puts these results in a
different perspective. The first set I obtained at significant computational
cost ($\sim n^{3}$) with another heuristic (``extremal optimization'',
EO) long ago in Ref. \citep{Boettcher03a} (black circles). The second
set is achieved by a simple gradient descent (GD, maroon squares).
GD sequentially looks at randomly selected (Boolean) variables $x_{i}$
among those whose flip ($x_{i}\mapsto\lnot x_{i}$) will improve the
cost function. (Such ``unstable'' variables are easy to track.)
After only $\sim0.4n$ such flips, typically no further improvements
were possible and GD converged; very scalable and fast (done overnight
on a laptop, averaging over $10^{3}-10^{5}$ instances at each $n$,
up to $n=10^{5}$). Presented in the form of Fig. \ref{fig:cutsize}(a),
the results \emph{all} look rather good, although it is already noticeable
that results for GD are barely distinguishable from those of the elaborate
GNN heuristic.

To discern further details, it is essential to present the data in
a form that, at least, eliminates some of its trivial aspects. For
example, as Schuetz et al reference themselves, the ratio $cut/n\sim\gamma$
converges to a stable limit with $\gamma\sim d/4+P_{*}\sqrt{d/4}+O(\sqrt{d})+o(n^{0})$
for $n,d\to\infty$ \citep{Dembo17}, where $P_{*}=0.7632\ldots$
\citep{parisi:80a}. In fact, for better comparison with Refs. \citep{Mezard03,Boettcher03a},
we focus on the average ground-state energy density of the Hamiltonian
in their Eq. (7) at $n=\infty$, which is related to $\gamma$ via
$\left\langle e_{d}\right\rangle /\sqrt{d}=\sqrt{d/4}-\gamma\sqrt{4/d}$.
(The awkward denominator is owed to fact that $P_{*}=\lim_{d\to\infty}\left\langle e_{d}\right\rangle /\sqrt{d}$.
Also, energy provides a fair reference point to assess relative error
because a purely random assignment of variables results in an energy
of zero, the ultimate null model. Such a reference point is lacking
for the errors quoted in Tab. 1 of Ref. \citep{Schuetz22}, for example.)

More revealing then merely dividing by $n$ is the transformation
of the data into an extrapolation plot \citep{Boettcher03a,Boettcher19}:
Since we care about the scalability of the algorithm in the asymptotic
limit for large problem sizes $n\to\infty$, which in the form of
Fig. \ref{fig:cutsize}(a) is out of view, it expedient to visualize
the data plotted for an \emph{inverse} of the problem size (i.e.,
$1/n$ or some power thereof \citep{Boettcher03a,Boettcher10b,Boettcher20}).
Independent of the largest sizes $n$ achieved in the data, it conveniently
condenses the asymptotic behavior arbitrarily close to the $y$-intercept
where $1/n\to0$, albeit it at the cost of sacrificing some data for
smaller $n$. To this end, I propose to plot the data in the finite-size
corrections form,
\begin{equation}
\left\langle e_{3}\right\rangle _{n}\sim\left\langle e_{3}\right\rangle _{n=\infty}+\frac{const}{n}+\ldots,\qquad(n\to\infty).\label{eq:FSC}
\end{equation}
In Fig. \ref{fig:cutsize}(b) we have plotted the same data from Fig.
\ref{fig:cutsize}(a) according to Eq. (\ref{eq:FSC}) for $d=3$
(modulo a trivial factor of $1/\sqrt{3}$ for better comparison with
$P_{*}$). Stark differences between each set of data appear, since
each set converges asymptotically to a stable but distinct limit at
$1/n=0$. First, we note the addition of a well-known result from
replica theory, a one-step replica symmetry-breaking (1-RSB) calculation
\citep{Mezard87,Mezard03} that is expected to yield the actual value
for $\left\langle e_{3}\right\rangle _{n=\infty}$ (and thus, $\gamma_{3}$)
with a precision of $10^{-4}$ (green line), a superior reference
value than $-P_{*}$ (black-dashed line), valid only at $d=\infty$
although seemingly sensible in the form of Fig. \ref{fig:cutsize}(a).
The 1-RSB value is further emphasized by the fact that the EO data
(black circles) from Ref. \citep{Boettcher03a} smoothly extrapolate
to the same limit within statistical errors. Finally, in the form
of Fig. \ref{fig:cutsize}(b), it becomes apparent that the claimed
GNN results (blue line) are systematical far ($>15\%$ at any $n$)
from optimal (1-RSB, green line) and hardly provide any improvement
over pure gradient descent (GD, maroon squares). It appears that the
GNN learns what is indeed the most typical about the energy landscape:
the vast prevalence of high-energy, poor-quality metastable solutions
that gradient descent gets trapped in, missing the faint signature
of exceedingly rare low-energy minima. In fact, extending GD by a
subsequent $5n$ spin flips, say, each flip adjusting one among the
least-stable spins (even if not always unstable), allows this greedy
local search to explore several local minima, still at linear cost.
The results of that simple algorithm, also shown in Fig. \ref{fig:cutsize}(b)
(diamonds), already reduce the error to $\approx6\%$ across all sizes
$n$, a considerable improvement on the GNN results in Ref. \citep{Schuetz22}
and still better than an improved version, GraphSAGE, the authors
mention in their response (orange line).

In conclusion, the study in Ref. \citep{Schuetz22} exemplifies a
number of common shortcomings found in the analysis of optimization
heuristics (see also Ref. \citep{Boettcher19}): (1) Reliance on rigorous
but rather poor and often meaningless bounds, as provided by the Goemans-Williamson
algorithm in this case, instead of using the much more relevant results
(albeit as-of-yet unproven) from statistical physics, (2) using an
obscure presentation of the data, (3) lack of state-of-the-art comparisons
across different areas in science, and (4) lack of benchmarking against
trivial, base-line models such as gradient descent or greedy search
we presented here. On such closer inspection, the proposed GNN heuristic
does not provide much algorithmic advantage over that base line. It
is likely that these conclusions are not isolated to this specific
example but would also hold for Max-Cut at $d=5$ and for the other
QUBO applications discussed in Ref. \citep{Schuetz22}, as the concurrent
comment by Angelini and Ricci-Tersenghi (arXiv:2206.13211) indicates. 

\bibliographystyle{apsrev4-1}
\bibliography{/Users/sboettc/Boettcher}

\end{document}